\documentclass[preprint2]{aastex}  

\shorttitle{Testing theories in barred spiral galaxies}
\shortauthors{Mart\'{\i}nez-Garc\'ia}

\begin{document}

\title{TESTING THEORIES IN BARRED SPIRAL GALAXIES}

\author{Eric E. Mart\'inez-Garc\'ia}
\affil{Instituto de Astronom\'ia, Universidad Nacional Aut\'onoma de M\'exico, AP 70-264, Distrito Federal 04510, Mexico.\\}
\email{martinez@astroscu.unam.mx}

\begin{abstract}

According to one version of the recently proposed ``manifold'' theory
that explains the origin of spirals and rings in relation to chaotic orbits,
galaxies with stronger bars should have a higher spiral arms pitch angle when
compared to galaxies with weaker bars.
A sub-sample of barred-spiral galaxies in the Ohio State University Bright Galaxy Survey,
was used to analyze the spiral arms pitch angle. These were compared with bar strengths taken from the literature.
It was found that the galaxies in which the spiral arms maintain a logarithmic
shape for more than 70$\degr$ seem to corroborate the predicted trend.

\end{abstract}


\section{INTRODUCTION}

Spiral arms in barred galaxies have been explained in the past as density waves~\citep[e.g.,][]{kor75}
or spiral waves that result from the crowding of gas orbits~\citep{hun78}.
\citet{kau96} invoked for the first time the need for chaotic orbits as
building blocks of spirals near the end of the bar. In the~\citet{kau96}~models, regular orbits dominate
the main structure of the bar and the outermost portions of spiral arms. The inner portions
of spiral arms are supported by chaotic orbits. 
Recently it has been proposed that chaotic motion can support the spirals in barred-spiral systems.
The new theory proposes that unstable Lagrangian
points ($L_{1}$ or $L_{2}$) near the end of the bar are the sites where chaotic orbits
are guided by invariant ``manifolds'', and are the origin of spirals
and (inner and outer) rings~\citep{vog06a,pat06,rom06,vog06b,vog06c,rom07,tso08,tso09,atha09a,har09,atha09b,atha10,con11}.
In this scenario the spiral dynamics are coupled to the bar,
and are driven by the manifolds.
This approach has been studied by two different groups of people. 

One of those groups~\citep{rom06,rom07,atha09a,atha09b,atha10},
considers a continuous flow of orbits along the manifolds emanating
from $L_{1}$ or $L_{2}$. When spirals form, stars move away from the corotation in a radial movement~\citep{atha10},
and material is needed to replenish the manifolds.
One prediction of this ``manifold theory'' (or ``Lyapunov tube model''),
not accounted for in the density wave scenario, is that stronger bars should have more open
spirals as compared to weaker bars, i.e., the spiral arms pitch angle\footnote{The angle between
a tangent to the spiral arm at a certain point and a circle,
whose center coincides with the galaxy's, crossing the same point.}
should increase with bar strength~\citep{atha09b}. This kind of
correlation was previously predicted by~\citet{sch84}, although for gas
arms driven by a bar perturbation.

Another view of the ``invariant manifold theory''~\citep{vog06b,vog06c,tso08,tso09}
considers the locus of all points with initial conditions at the unstable manifolds
that reach a local apocentric~\citep[or pericentric, see][]{har11} passage, i.e., the apsidal
sections of the manifolds. In this scenario, there is no need for the replenishment of material
to obtain long-lived spirals~\citep[see, e.g.,][]{eft10}. 
Both views of the ``invariant manifold theory'' predict a trailing spiral pattern
for strong perturbations and similar pattern speeds for the bar and spiral, i.e.,
$\Omega_{p}^{\mathrm{bar}}=\Omega_{p}^{\mathrm{spiral}}$.
However, in the view of~\citet{vog06b,vog06c} and~\citet{tso08,tso09}, the ``azimuthal tilt'' of the spiral
response~\citep{tso09},
i.e., the difference between the bar's major axis and the Lagrangian points $L_{1}$ or $L_{2}$ at the
moment of the onset of the spiral, determines how open the spiral arms will be.
In this case, the pitch angles are smaller than the ones predicted by~\citet{atha09b}
and become even smaller for pure bar models when the ``azimuthal tilt'' is not taken
into account (C. Efthymiopoulos, private communication 2011).

\citet{pat10} describe one more dynamical mechanism
that supports spiral arms through stars in chaotic motion.
They propose this mechanism by describing the spiral arms of the barred-spiral NGC 1300.
Together with the bar, these spiral arms are inside the corotation and are not related
to the presence of unstable Lagrangian points and the associated families of periodic orbits.
This alternative mechanism may be linked to some range of pitch angles of spiral arms
encountered in barred-spiral systems.

Do manifolds drive spiral dynamics in barred galaxies?
Or are the dynamics driven by the bar?
The bar may drive the dynamics, affecting the spiral amplitude locally,
as reported by~\citet{sal10}~\citep[see also][]{blo04} and previously discarded (or weakly corroborated)
by other authors comparing bar strength to spiral arm strength~\citep{but09,dur09,sei98}.
Bars driving the dynamics would imply an accordance with
(linear) density wave theory. These spirals may be a continuation of the
bar mode, or an independent mode coupled to the bar~\citep[e.g.,][]{tag87,mass97}.
In the ``Lyapunov tube model'', the strength of the bar affects the pitch angle of the spirals,
but not its amplitude. The amplitude of the spirals depends on how much material is trapped by
the manifolds, although, the amplitude of the spirals should in general decrease outward~\citep{atha10}.
~\citet{gros04} investigated the relation between the amplitude of the spirals with the
pitch angle in non-barred and weakly barred galaxies.

One prediction of the density wave theory~\citep[][see~\S\ref{denw_pred}]{hoz03}
entails that different pitch angles are expected for spirals when
observed in different bands (e.g., optical versus near-infrared [NIR]).
According to~\citet{atha10}, the ``invariant manifold theory'' predicts that stars of different
ages will be guided by the same manifold, and no difference
between the winding of the spirals is expected.

In this paper, we will investigate 
whether the predictions of pitch angles are observed for real galaxies, or not.
Two methods were applied for this purpose: the ``slope method'' (Section~\ref{sec_slope}),
which is especially good for determining how long the logarithmic shape
is maintained for spiral arms, and the ``Fourier method'' (Section~\ref{sec_fourier}),
which was used to determine the ``dominant'' pitch angle inside a given annulus
for each object.

\section{GALAXY SAMPLE}~\label{gal_sample}

The initial galaxy sample consists of 104 galaxies classified as Fourier bars
in~\citet{lau04}. The data were acquired from the Ohio State University 
Bright Galaxy Survey (OSUBGS)~\citep{esk02}. From this initial sample,
it was found that only 84 objects present spiral-like features. Nevertheless,
not all the objects are suitable for this kind of study due to asymmetries, e.g.,
short, faint, or ragged spiral arms, or prominent rings. The following criteria
were established in order to obtain a sample, including objects with a morphology
candidate to be explained by ``chaotic'' spirals.

\begin{enumerate}

\item The spiral arms must remain logarithmic, i.e., with a constant pitch angle ($i$),
at least for $50\degr$ in the azimuthal range, $\alpha$.\footnote{Although the spiral arms may
extend further in the disk with a varying pitch angle, i.e., different slopes in 
a $\ln{r}$ versus $\theta$ map.}
This was verified with the ``slope method'' (see~Section~\ref{sec_slope}).
The lower limit value of $\alpha$ was chosen according to Figure 4 in~\citet{atha09b}, where the manifold loci remain
logarithmic (for the adopted model parameters) and maintain a ``nearly'' logarithmic geometry
up to $\sim 100\degr$. We consider that the manifold loci and the density maximum along the spirals coincide.
According to~\citet{pat06}, spirals supported by chaotic
particles may extend up to $\pi/2$ radians.
Variations of $\alpha$ toward larger angles will be discussed in~Section~\ref{sec_resdis}.  

\item The object presents two spiral arms visually connected to the bar.

\item No prominent inner rings (near the bar's end) are present.\footnote{An exception
is NGC 5921 where a ring is present, but it does not dominate over the spiral features.}
Ring structures are connected to the bar on both sides. The pitch angle definition
as applied in this investigation only refers to spiral arms.
A dependence of the inner ring shape on bar strength has been investigated by~\citet{gro10}.

\end{enumerate}

\noindent After applying these selection criteria, the final sample consists of 27 barred
spirals (see, e.g., Table~\ref{tbl-Qs}).

In order to use the bar strength values
of~\citet[][see~Section~\ref{bar_strength}]{lau04}, we adopt the same deprojection
parameters of those authors, i.e. the same values for position angle ($\phi$) and minor-to-major
axial ratios ($q=b/a$).\footnote{With the exception of NGC 1300,
for which we adopt $\phi = 100 \degr \pm 14$, and $q= 0.6 \pm 0.1$ (see~Section~\ref{sec_OBJcomments}).
Uncertainties of 10$\%$ in inclination translate in 10$\%$-15$\%$ in
perturbation strength~\citep{but01,lau02}. }
To determine these parameters,~\citet{lau04} fit ellipses to the outer isophotes on the disk.
They were based on the OSUBGS $B$-band images that are deeper than $H$-band images.

To test the~\citet{atha09b,atha10} predictions regarding spiral arms pitch angles,
we use the NIR $H$-band since we are interested
in ``long''-lived structures rather than young stars, HII regions, or gas that would be present in optical data.

\vspace{10 mm}

\section{BAR STRENGTH}~\label{bar_strength}

The predicted trend in~Athanassoula et al.'s (2009a) ``manifold models'' requires the
strength of the bar at the radius of the Lagrangian points $L_{1}$ or $L_{2}$.
It should be mentioned that for these models the self-gravity of the spirals was
not taken into account. On the other hand, the addition of the spiral potential
in~Tsoutsis et al.'s (2009) models shifts the positions of the Lagrangian points $L_{1}$ or $L_{2}$
both in the radial and azimuthal directions.

The strength of the bar can be obtained from the~\citet{lau04} radial profiles of the perturbation
strength.~\citet{lau04} used the gravitational torque method~\citep{com81,but01,blo02}
taking care of the artificial bulge stretch~\citep[see also][]{spe08}.
The perturbation strength is calculated as

\begin{equation}
   Q_{t}(r) = \frac{~~\left( \frac{\partial{\Phi(r,\theta)}}{ \partial{\theta}} \right)_\mathrm{max} }
              {r \frac{\mathrm{d}{\Phi_{0}(r)}}{\mathrm{d}{r}}},     
\end{equation}
          
\noindent which represents the ratio between
the maximum amplitude (over azimuth) of the tangential
force, and the mean axisymmetric radial force
derived from the $m=0$ component of the gravitational potential.
The potential is inferred from the luminous mass,
and can be represented as~\citep[see][]{com81,qui94}:

\begin{equation}
         \Phi(r,\theta) \approx \Phi_{0}(r) + \sum_{m=2,4,6} \Phi_{m}(r) \cos{[m\theta]}.
\end{equation}

\noindent The angle $\theta$ is given in the deprojected image,
and $\theta=0$ along the bar major axis.
For this investigation we assume that $L_{1}=L_{2}=L$. For real galaxies, $L_{1}$ may differ
from $L_{2}$ due to odd terms in the gravitational potential.

We analyzed three cases in which the bar's strength is estimated in three different ways.

\begin{enumerate}

 \item In the first case, the bar's strength is estimated at $r=r_{L}$.
The Lagrangian point or corotation radius~\citep{sell93}, $r_{L}$, was obtained from~\citet{butz09}, who
applied the ``potential-density phase shift method'' to the OSUBGS sample.
There have been significant discussions on the validity of this method. This is partly 
because~\citet{zha07} found some cases (e.g., NGC 4665) where
$r_{L}/r_{\mathrm{bar}} < 1$, i.e., corotation before the end of the bar. 
According to~\citet{con80}, self-consistent bars are not possible to be modeled in this regime.

One important difference between ``manifold'' models~\citep{rom06,vog06b,vog06c,atha09a} and 
the ``potential-density phase shift method'' is that~\citet{zha07} and~\citet{butz09} considered potentials
varying considerably with time.
The time-independent (rigid) potentials of the ``manifold'' models generate ``passive'' chaotic orbit responses.
Although the~\citet{zha07} models involve chaos in the individual stars' trajectories, ``collective dissipation'' makes
possible the existence of coherent structures (e.g., spiral arms).
We define ``$r_{\mathrm{BZ09}}$'' as the corotation radius obtained from~\citet{butz09}.
Table~\ref{tbl-Qs} shows the $Q_{t}(r=r_{\mathrm{BZ09}})$ values for the 27 OSUBGS barred galaxies.

 \item In the second case, we estimate the bar's strength at a distance $r_{L}=1.2r_{\mathrm{bar}}$.
According to various studies~\citep{ath92,elm96,ague03}, the expected range for the bar length
lies between $r_{\mathrm{bar}}=r_{L}/1.0$ and $r_{\mathrm{bar}}=r_{L}/1.4$.
~\citet{elm96} and~\citet{ague03} also discuss objects where $r_{\mathrm{bar}}=r_{L}/1.7$.
A mean value of $r_{\mathrm{bar}}=r_{L}/1.2$
is expected for large samples of galaxies. 
For bar strengths, the effect of having $r_L=1.0r_{\mathrm{bar}}$ or $r_L=1.4r_{\mathrm{bar}}$,
instead of $r_L=1.2r_{\mathrm{bar}}$,
could be much larger than deprojecting a galaxy within $10\%$ error in the projection angles.
For this study the bar length, $r_{\mathrm{bar}}$, was
taken from~\citet{lau04}. In Table~\ref{tbl-Qs} we show the $Q_{t}(r=1.2r_{\mathrm{bar}})$ values
adopted for this investigation.

 \item The third case involves the maximum of the radial $Q_{t}(r)$ profiles
or $Q_{g}$. These were tabulated in~\citet{lau04}.

\end{enumerate}

The adopted $Q_{t}(r)$ values from~\citet{lau04} were computed assuming a
constant $M/L$ ratio throughout the disk and an empirical correlation
for the vertical scale-height ($h_{z}$). Also, it is assumed that
dark matter has little impact on the bar strength.
For the $Q_{t}(r)$ error calculation shown in Table~\ref{tbl-Qs},
the $Q_{g}$ error of~\citet{lau04} was summed in quadrature with
the error inherent to digitization\footnote{The
ADS's data extraction applet,~{\it{DEXTER}}~\citep{dem01}, was used for this purpose.}
of the $Q_{t}(r)$ plots and the $r_{L}$~\citep{butz09} errors for
the $Q_{t}(r=r_{\mathrm{BZ09}})$ values.

A technique for separating the gravitational torques of bars and spirals 
was developed by~\citet{but03,but05}. This technique separates the bar+disk
image to obtain the bar strength $Q_{b}$ (at the
respective maximum of $Q_{t}(r)$) unaffected by the spiral
gravitational influence.
Nevertheless, for the majority
of barred galaxies in the OSUBGS sample, the bar strength, $Q_{b}$,
dominates over the spiral arm strength $Q_{s}$~\citep{dur09}.
Also, the correction of the spiral arms does not affect the tendencies
for $Q_{g}$ in the Hubble sequence~\citep{lau07}.
In either case, for this investigation it is assumed that $Q_{g} \sim Q_{b}$,
and that the $Q_{t}(r)$ values are affected by the spirals within the errors.

\section{PITCH ANGLES}~\label{sec_pitch}

Spiral arms pitch angles have been measured in the literature
with different methods.~\citet{dan42} measured the spiral arms
on photographic plates.~\citet{ken81} measured the spiral shapes using
the intensity and HII region distributions.~\citet{ma99}
fit the shapes of spiral arms directly on the images.
Fourier decomposition methods had also been used~\citep[e.g.,][]{con88,pue92,sch94,sei06},
yielding similar results as other methods~\citep{con88,pue92}.


\begin{figure*}
\centering
\includegraphics[scale=0.70]{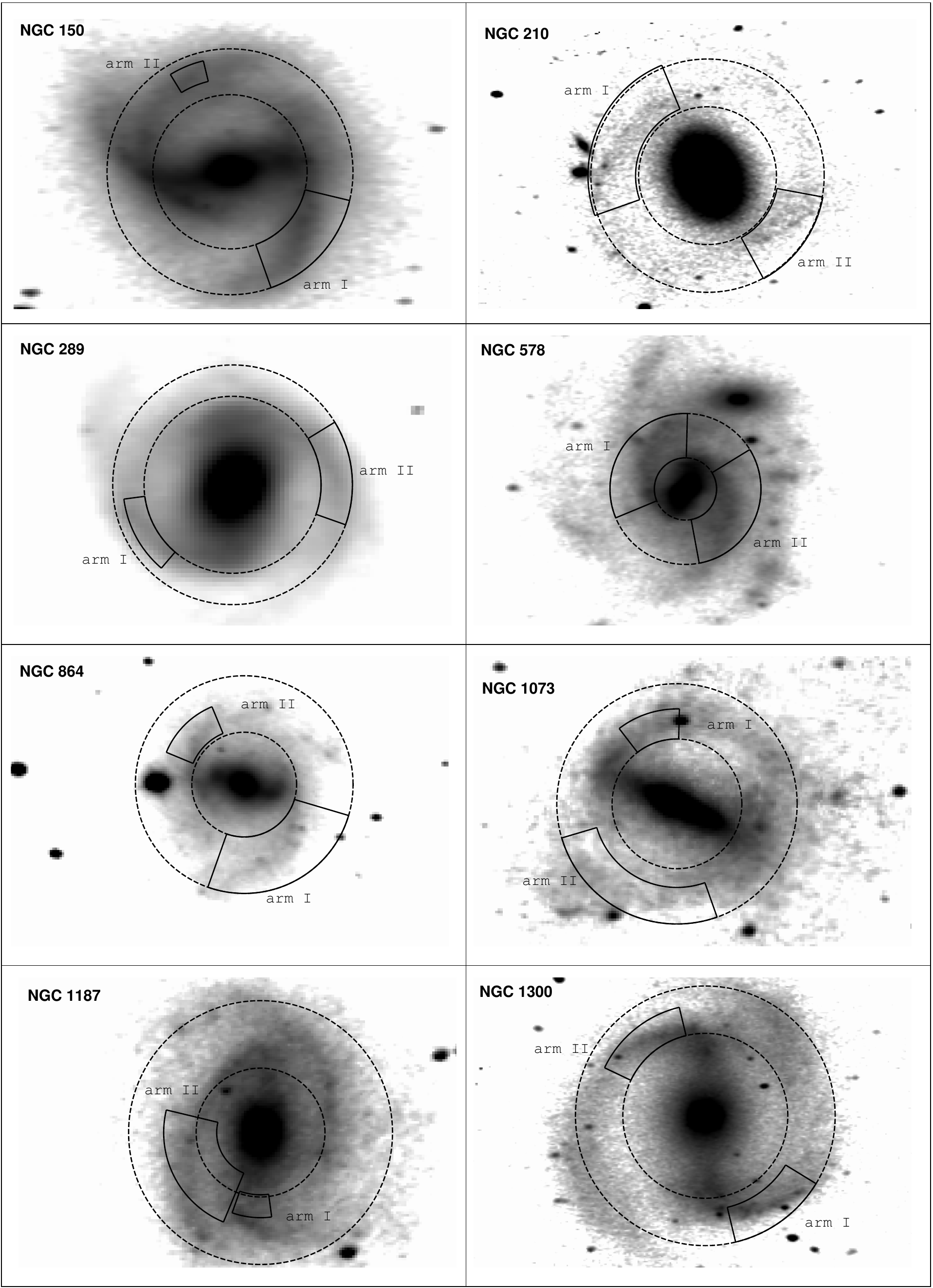}
 \caption{Deprojected $H$-band images of NGC 150, NGC 210, NGC 289, NGC 578, NGC 864, NGC 1073, NGC 1187, and NGC 1300.
          The display is in a logarithmic scale. The analyzed arm segments for the ``slope method'' are shown in the figures (solid lines),
          together with the annulus adopted for the ``Fourier method'' (dashed lines).}
 \label{combo_1}
\end{figure*}

\begin{figure*}
\centering
\includegraphics[scale=0.70]{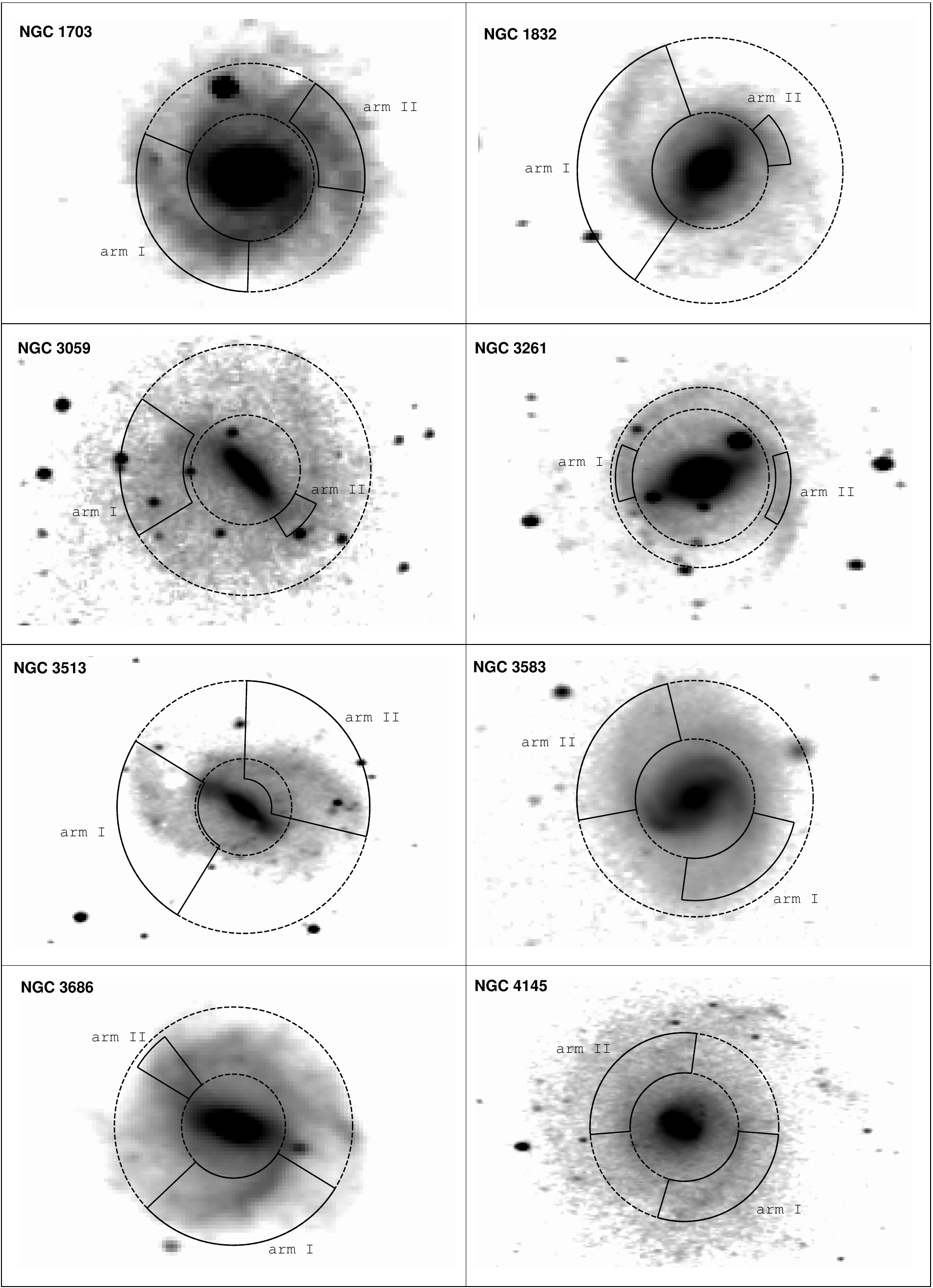}
 \caption{Deprojected $H$-band images of NGC 1703, NGC 1832, NGC 3059, NGC 3261, NGC 3513, NGC 3583, NGC 3686, and NGC 4145.
          The display is in a logarithmic scale.}
 \label{combo_2}
\end{figure*}

\begin{figure*}
\centering
\includegraphics[scale=0.70]{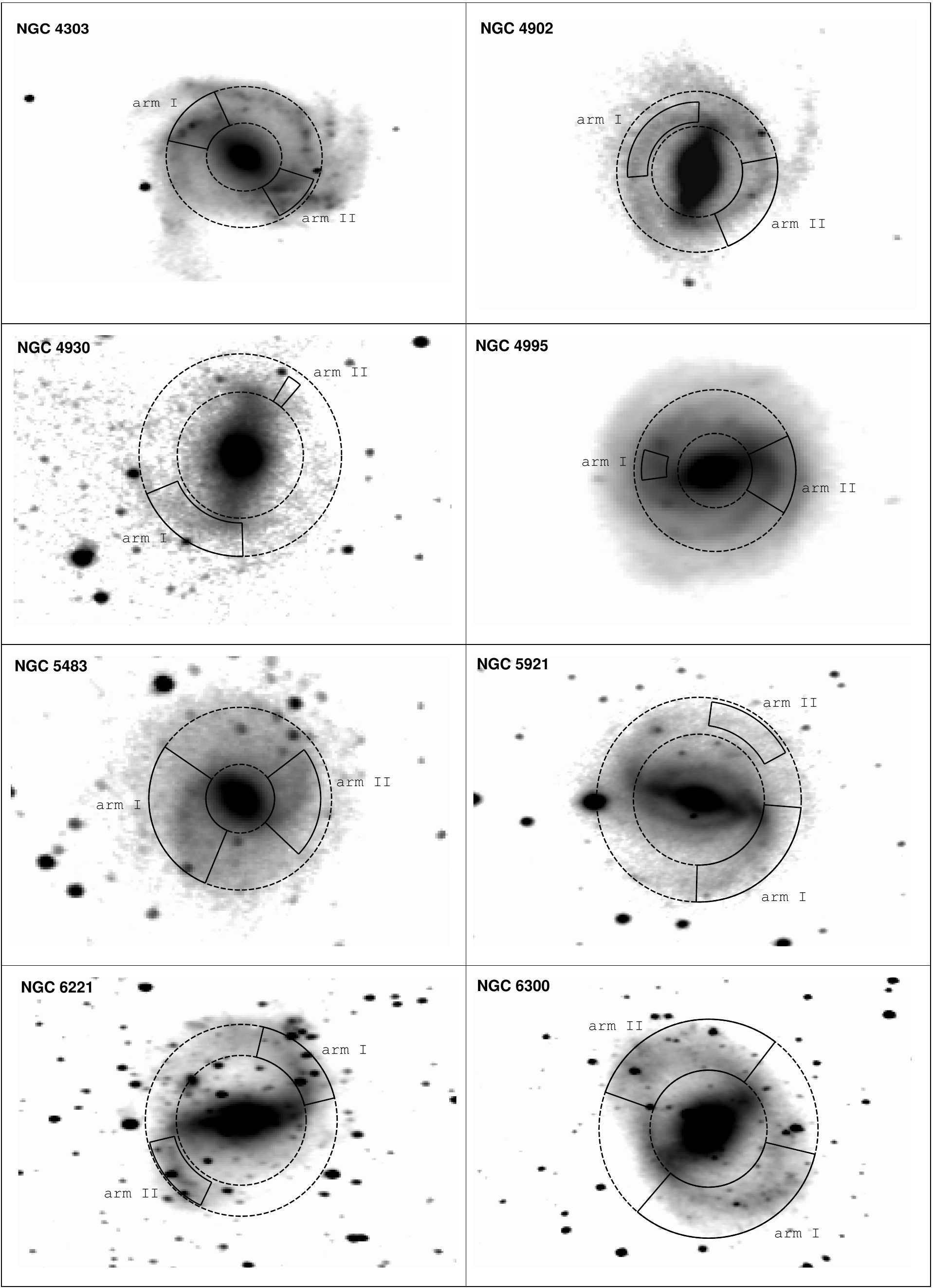}
 \caption{Deprojected $H$-band images of NGC 4303, NGC 4902, NGC 4930, NGC 4995, NGC 5483, NGC 5921, NGC 6221, and NGC 6300.
          The display is in a logarithmic scale.}
 \label{combo_3}
\end{figure*}

\begin{figure*}
\centering
\includegraphics[scale=0.70]{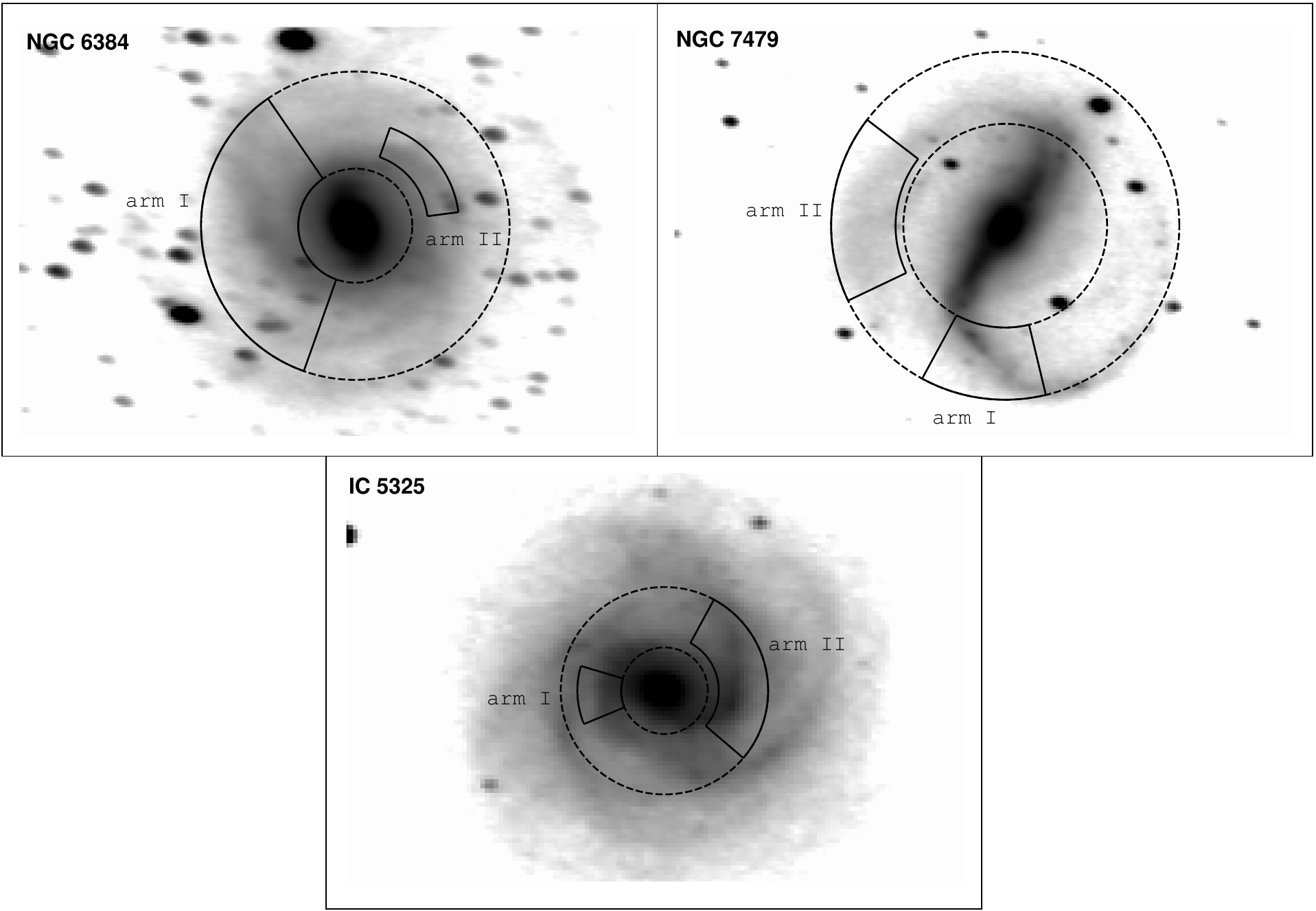}
 \caption{Deprojected $H$-band images of NGC 6384, NGC 7479, and IC 5325.
          The display is in a logarithmic scale.}
 \label{combo_4}
\end{figure*}


\begin{deluxetable}{cllcll}
\tabletypesize{\scriptsize}
\tablecaption{Perturbation Strengths~\label{tbl-Qs}}
\tablewidth{0pt}
\tablehead{
\colhead{Galaxy} & \colhead{$Q_{t}(r=r_{\mathrm{BZ09}})$} & \colhead{$Q_{t}(r=1.2r_{\mathrm{bar}})$} &
\colhead{Galaxy} & \colhead{$Q_{t}(r=r_{\mathrm{BZ09}})$} & \colhead{$Q_{t}(r=1.2r_{\mathrm{bar}})$}
}

\startdata

NGC 150      & 0.302   $\pm 28.1\%$  & 0.352   $\pm 24.1\%$  & NGC 3686     & 0.075   $\pm 24.0\%$  & 0.152   $\pm 11.8 \%$ \\
NGC 210      & 0.060   $\pm  5.0\%$  & 0.058   $\pm  3.4\%$  & NGC 4145     & 0.119   $\pm  2.5\%$  & 0.122   $\pm  2.5 \%$ \\
NGC 289      & 0.133   $\pm  3.8\%$  & 0.099   $\pm  7.1\%$  & NGC 4303     & 0.149   $\pm 29.5\%$  & 0.251   $\pm 17.5 \%$ \\
NGC 578      & 0.092   $\pm 79.3\%$  & 0.040   $\pm 37.5\%$  & NGC 4902     & 0.068   $\pm 38.2\%$  & 0.128   $\pm 20.3 \%$ \\
NGC 864      & 0.201   $\pm 18.9\%$  & 0.236   $\pm 15.7\%$  & NGC 4930     & 0.138   $\pm 15.9\%$  & 0.099   $\pm 22.2 \%$ \\
NGC 1073     & 0.498   $\pm  6.4\%$  & 0.386   $\pm  3.4\%$  & NGC 4995     & 0.280   $\pm 16.8\%$  & 0.263   $\pm 17.9 \%$ \\
NGC 1187     & 0.162   $\pm 26.5\%$  & 0.206   $\pm 20.9\%$  & NGC 5483     & 0.098   $\pm  4.1\%$  & 0.111   $\pm  3.6 \%$ \\
NGC 1300     & 0.475   $\pm  3.4\%$  & 0.277   $\pm  4.0\%$  & NGC 5921     & 0.384   $\pm  6.0\%$  & 0.329   $\pm  7.0 \%$ \\
NGC 1703     & 0.100   $\pm  5.0\%$  & 0.058   $\pm  8.6\%$  & NGC 6221     & 0.309   $\pm 36.2\%$  & 0.237   $\pm 47.3 \%$ \\
NGC 1832     & 0.121   $\pm 19.8\%$  & 0.149   $\pm 16.1\%$  & NGC 6300     & 0.158   $\pm  6.3\%$  & 0.080   $\pm  3.8 \%$ \\
NGC 3059     & 0.251   $\pm 20.3\%$  & 0.268   $\pm 17.9\%$  & NGC 6384     & 0.136   $\pm 14.7\%$  & 0.048   $\pm 41.7 \%$ \\
NGC 3261     & 0.133   $\pm  6.8\%$  & 0.117   $\pm  7.7\%$  & NGC 7479     & 0.516   $\pm 11.6\%$  & 0.240   $\pm 25.0 \%$ \\
NGC 3513     & 0.256   $\pm 27.0\%$  & 0.210   $\pm 32.9\%$  &  IC 5325     & 0.211   $\pm  9.5\%$  & 0.102   $\pm 19.6 \%$ \\
NGC 3583     & 0.160   $\pm  5.0\%$  & 0.207   $\pm  2.9\%$  & ~            & ~                     & ~                     \\

\enddata

\tablecomments{
Columns (1) and (4): object name. 
Columns (2) and (5): perturbation strength (see Section~\ref{bar_strength})
from~\citet{lau04}, at corotation radius from~\citet{butz09}.
Columns (3) and (6): perturbation strength from~\citet{lau04}, at radius $r=1.2r_{\mathrm{bar}}$
(see Section~\ref{bar_strength}).
}

\end{deluxetable}


\subsection{``Slope Method''}~\label{sec_slope}

This method is similar to the one
used in~\citet{sei98}. It is assumed that the arms can be
represented by logarithmic spirals, which implies a constant pitch angle.
Although, variable pitch angles may be a better and more adequate 
representation for some objects~\citep[see e.g., the case of NGC 1365 in][]{rin09}.

Before deprojection, the spiral regions were isolated by masking the bar,
foreground stars, strong star-forming regions (visually selected), bad pixels,
and other structures not associated with the corresponding arm region.
After deprojection ($H$-band data), the centers of the objects were
determined by fitting ellipses to the central isophotes close to the
bar region.\footnote{This was done with the ELLIPSE task in IRAF.
IRAF is distributed by the National Optical Astronomy Observatory,
which is operated by the Association of Universities for Research in
Astronomy, Inc., under cooperative agreement with the National Science
Foundation.}
Afterward, the spiral arms
were ``unwrapped'' by plotting them in a $\ln{r}$ versus $\theta$
map~\citep[e.g.,][]{iye82,elm92,gros04}.
Under this geometric transformation, logarithmic spirals appear as straight
lines. The pitch angle, $i$, is related to the slope of the
line, $s$, as

\begin{equation}
 \cot{i} = k |s|,
\end{equation}

\noindent wherein $k$ is a constant\footnote{Basically converts $\theta$-axis pixels to radians, and
determines the equivalence between pixels in the $\ln{r}$ axis and physical units of an image.}
due to ``pixelation'' and unit conversion.

\begin{figure}
\centering
\epsscale{1.05}
\plotone{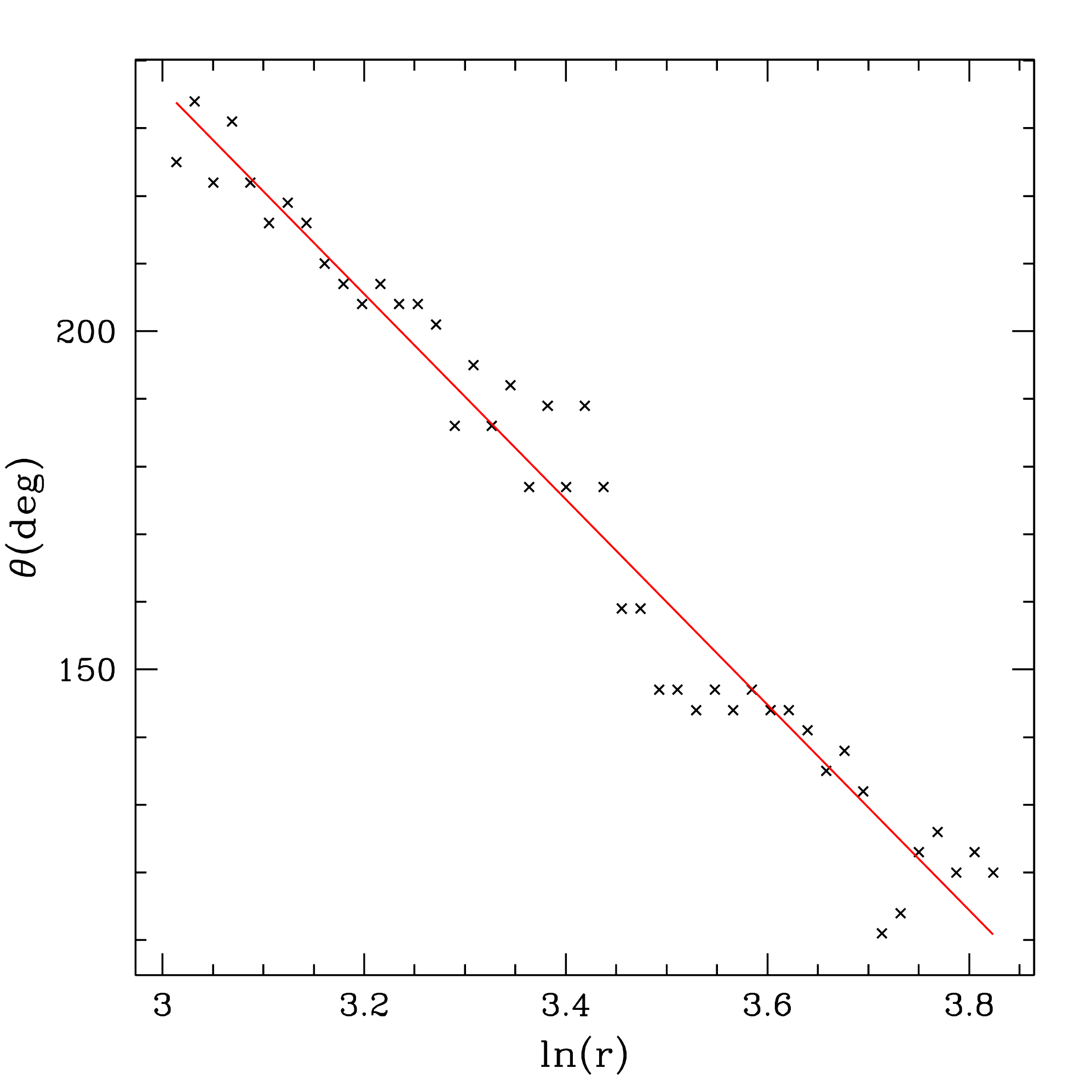}
 \caption{Plot of $\ln{r}$ vs. $\theta$ for arm segment I in NGC 1832 ($H$-band).
Crosses indicate the points where a maximum intensity was
found for each column for the corresponding section in the ``unwrapped'' image. 
 The continuous line indicates the least-squares fit.}
 \label{fig_slope}
\end{figure}

Two arm segments were selected closest to the bar's end with the
condition that the slope, $s$, was maintained nearly constant along them
(see Figures~\ref{combo_1}-\ref{combo_4}).
Due to this ``slope restriction'', in many cases the critical segment including the part of the arms attached
to the bar, was not able to be considered. 
The slope ($s$) is determined by first selecting for each
column in the arm segment the pixels with a maximum in intensity
(see as an example Figure~\ref{fig_slope}, for the case of NGC~1832).
A least-squares fit is then obtained for the resulting pixels.
As already mentioned, these fits were done in the $H$-band aiming to trace Population II stars.
Young stars and clusters can contribute locally up to $20\%$-$30\%$ of the observed radiation
in the NIR~\citep[e.g.,][]{rix93,rho98,pat01,gros08}. How these young objects affect
the pitch angles' measurements depends on the star formation conditions and young stars kinematics.
For this investigation, it is assumed that young stars and clusters affect the spiral arms pitch angles
within the errors involved in the methods applied.

As previously mentioned, all the objects with inner rings,
asymmetries, unclear, or ``logarithmically short'' arms were discarded
from the analysis. For the remaining 27 objects, the arm segments (I or II)
best determined and with the clearest spiral structure were also identified.
These are marked with an asterisk~(*) in Table~\ref{tbl-slope}, together with the adopted radial ranges,
$\Delta{r}$, tabulated from innermost ($r_{0}$) to outermost radius ($r$), and azimuthal ranges, $\alpha$.
Azimuthal ranges are obtained by the equation

\begin{equation}~\label{eq_alpha}
 \alpha = \cot{i_{H}} \ln{\left( \frac{r}{r_{0}} \right)}
\end{equation}

\noindent and are displayed graphically in Figures~\ref{combo_1}-\ref{combo_4} with regions delimited by solid lines.
These values do not indicate the end of the spirals, since spiral arms may extend further with
a variable pitch angle~\citep{rin09}. Although if the extensions of the spirals have reduced amplitudes with
respect to the logarithmic part, their phases will be difficult to determine, as will their pitch angles.
The estimations are done independently of the amplitude (strength) of the spiral itself.
~\citet[][their Figure 8]{gros04} found a tendency between the amplitude of the $m=2$ spiral and pitch angles
in SA and SAB galaxies. 

\begin{figure*}
\centering
\epsscale{1.5}
\plotone{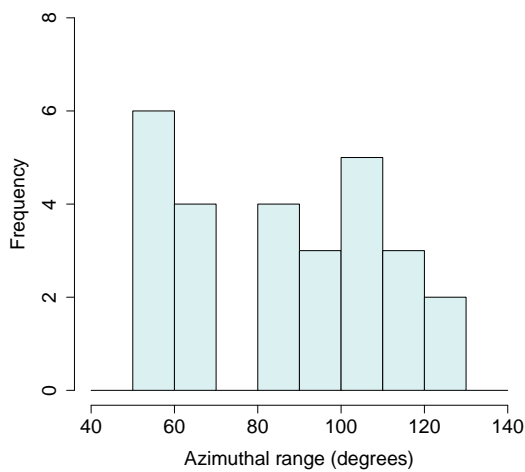}
 \caption{Histogram of the maximum azimuthal ranges (either arm segment I or II, see Table~\ref{tbl-slope}).}
\label{histo_II}
\end{figure*}

Figure~\ref{histo_II} shows a histogram of the maximum azimuthal range distribution (either arm segment I or II)
for each object presented in Table~\ref{tbl-slope}.

\subsubsection{Error Determination}
Errors introduced by deprojection parameters ($\phi$ and $q$)
translate into different slopes or deviations of a straight line
in the $\ln{r}$ versus $\theta$ plots.
For each object, five deprojected frames were obtained to better
account for these errors. The images were deprojected with the parameters 
$\phi, q$; $\phi+sd, q$; $\phi-sd, q$; $\phi, q+sd$;
and $\phi, q-sd$, where $sd$ is the respective standard deviation.
Pitch angle values were measured and compared to the case when $\phi$ and $q$
were used as the deprojection parameters (i.e., when $sd=0$). The cases with the highest (positive)
or lowest (negative) discrepancies were adopted to account for the $+\sigma$
and $-\sigma$ errors, respectively (see Table~\ref{tbl-slope}).


\begin{deluxetable}{llccllcc}
\tabletypesize{\scriptsize}
\tablecaption{``Slope Method" Derived Parameters~\label{tbl-slope}}
\tablewidth{0pt}
\tablehead{
\colhead{Galaxy and Segment} & \colhead{$i_{H}$(deg)} & \colhead{$\Delta{r}$(arcsec)} & \colhead{$\alpha$(deg)} &
\colhead{           Segment} & \colhead{$i_{H}$(deg)} & \colhead{$\Delta{r}$(arcsec)} & \colhead{$\alpha$(deg)}
}

\startdata

\vspace{1 mm}
NGC 150  Arm I*  & $ 24.6^{+ 5.3}_{-1.5}$ & (34.7-55.4)  & $ 58 \pm  8  $ &         Arm II  & $ 33.2^{+ 7.6}_{-10.3}$ & (41.8-51.4) & $ 18 \pm  5 $ \\
\vspace{1 mm}
NGC 210  Arm I*  & $ 18.4^{+ 0.7}_{-0.9}$ & (64.8-107.6) & $ 88 \pm  4  $ &         Arm II  & $ 29.5^{+ 1.8}_{-1.8}$  & (63.6-105.6) & $ 51 \pm  4 $ \\
\vspace{1 mm}
NGC 289  Arm I   & $ 16.5^{+ 2.1}_{-1.2}$ & (22.8-28.2)  & $ 41 \pm  4  $ &         Arm II* & $ 19.3^{+ 4.8}_{-0.9}$  & (22.8-30.8) & $ 50 \pm  7 $ \\
\vspace{1 mm}
NGC 578  Arm I*  & $ 23.8^{+ 0.4}_{-1.3}$ & (20.8-50.3)  & $115 \pm  4  $ &         Arm II  & $ 24.8^{+ 1.4}_{-1.3}$  & (20.8-50.3) & $110 \pm  7 $ \\
\vspace{1 mm}
NGC 864  Arm I*  & $ 24.3^{+ 0.7}_{-0.6}$ & (33.0-68.5)  & $ 93 \pm  3  $ &         Arm II  & $ 28.8^{+ 4.1}_{-1.2}$  & (34.8-53.1) & $ 44 \pm  4 $ \\
\vspace{1 mm}
NGC 1073 Arm I*  & $ 29.0^{+ 7.8}_{-6.7}$ & (37.8-55.3)  & $ 39 \pm 10  $ &         Arm II  & $ 12.6^{+ 4.4}_{-1.8}$  & (48.7-70.0) & $ 93 \pm 18 $ \\
\vspace{1 mm}
NGC 1187 Arm I   & $ 35.7^{+ 3.8}_{-4.1}$ & (35.7-48.8)  & $ 25 \pm  3  $ &         Arm II* & $ 29.1^{+ 3.2}_{-0.5}$  & (25.2-55.6) & $ 82 \pm  6 $ \\
\vspace{1 mm}
NGC 1300 Arm I*  & $ 21.5^{+ 7.4}_{-7.8}$ & (84.8-116.9) & $ 47 \pm 14  $ &         Arm II  & $ 18.8^{+ 6.1}_{-5.5}$  & (74.3-100.5) & $ 51 \pm 13 $ \\
\vspace{1 mm}
NGC 1703 Arm I*  & $ 16.7^{+ 0.4}_{-1.9}$ & (15.6-28.1)  & $112 \pm  7  $ &         Arm II  & $ 25.0^{+ 0.7}_{-0.7}$  & (16.8-28.1) & $ 64 \pm  2 $ \\
\vspace{1 mm}
NGC 1832 Arm I*  & $ 20.7^{+ 1.3}_{-1.2}$ & (20.0-45.8)  & $126 \pm  8  $ &         Arm II  & $ 27.0^{+ 6.1}_{-2.9}$  & (20.0-27.8) & $ 37 \pm  6 $ \\
\vspace{1 mm}
NGC 3059 Arm I*  & $ 31.5^{+ 1.9}_{-1.6}$ & (26.8-54.0)  & $ 66 \pm  4  $ &         Arm II  & $ 33.7^{+ 7.5}_{-2.1}$  & (23.6-33.9) & $ 31 \pm  5 $ \\
\vspace{1 mm}
NGC 3261 Arm I*  & $ 17.1^{+15.6}_{-2.5}$ & (27.4-34.2)  & $ 41 \pm 14  $ &         Arm II  & $ 11.8^{+ 3.3}_{-0.8}$  & (30.0-36.1) & $ 50 \pm  7 $ \\
\vspace{1 mm}
NGC 3513 Arm I   & $ 32.9^{+ 1.3}_{-0.9}$ & (26.1-72.6)  & $ 91 \pm  4  $ &         Arm II* & $ 40.7^{+ 1.4}_{-1.7}$  & (16.3-72.6) & $100 \pm  5 $ \\
\vspace{1 mm}
NGC 3583 Arm I   & $ 20.4^{+ 0.5}_{-3.3}$ & (34.7-59.3)  & $ 83 \pm  7  $ &         Arm II* & $ 24.1^{+ 2.0}_{-0.7}$  & (34.7-68.8) & $ 87 \pm  5 $ \\
\vspace{1 mm}
NGC 3686 Arm I*  & $ 24.6^{+ 0.9}_{-0.5}$ & (20.2-46.2)  & $104 \pm  3  $ &         Arm II  & $ 67.0^{+ 6.6}_{-2.2}$  & (20.2-43.7) & $ 19 \pm  4 $ \\
\vspace{1 mm}
NGC 4145 Arm I*  & $ 17.0^{+ 2.4}_{-0.4}$ & (50.5-87.9)  & $104 \pm  8  $ &         Arm II  & $ 17.6^{+ 3.3}_{-1.4}$  & (50.5-87.9) & $100 \pm 12 $ \\
\vspace{1 mm}
NGC 4303 Arm I*  & $ 38.4^{+ 1.0}_{-1.2}$ & (29.2-60.5)  & $ 53 \pm  2  $ &         Arm II  & $ 44.1^{+ 1.5}_{-1.9}$  & (29.2-57.3) & $ 40 \pm  2 $ \\
\vspace{1 mm}
NGC 4902 Arm I*  & $ 11.2^{+ 6.6}_{-1.8}$ & (21.4-30.0)  & $ 98 \pm 26  $ &         Arm II  & $ 23.1^{+ 5.3}_{-0.6}$  & (19.5-34.7) & $ 77 \pm  9 $ \\
\vspace{1 mm}
NGC 4930 Arm I*  & $ 18.7^{+ 0.6}_{-2.7}$ & (38.9-58.2)  & $ 68 \pm  6  $ &         Arm II  & $ 74.6^{+ 5.5}_{-16.1}$ & (36.2-53.1) & $  6 \pm  4 $ \\
\vspace{1 mm}
NGC 4995 Arm I   & $ 47.8^{+ 3.7}_{-4.9}$ & (22.6-34.3)  & $ 22 \pm  3  $ &         Arm II* & $ 38.5^{+ 6.3}_{-2.1}$  & (17.3-37.7) & $ 56 \pm  8 $ \\
\vspace{1 mm}
NGC 5483 Arm I*  & $ 29.2^{+ 0.5}_{-4.4}$ & (14.8-39.4)  & $100 \pm  9  $ &         Arm II  & $ 31.8^{+ 1.4}_{-3.0}$  & (14.8-34.7) & $ 79 \pm  6 $ \\
\vspace{1 mm}
NGC 5921 Arm I*  & $ 16.1^{+ 0.5}_{-0.6}$ & (61.0-95.3)  & $ 89 \pm  3  $ &         Arm II  & $ 16.4^{+ 0.9}_{-1.1}$  & (69.5-91.9) & $ 54 \pm  3 $ \\
\vspace{1 mm}
NGC 6221 Arm I*  & $ 19.5^{+ 1.3}_{-0.7}$ & (39.0-57.6)  & $ 63 \pm  3  $ &         Arm II  & $ 20.0^{+ 2.5}_{-1.0}$  & (41.2-56.5) & $ 50 \pm  4 $ \\
\vspace{1 mm}
NGC 6300 Arm I*  & $ 17.3^{+ 0.6}_{-0.4}$ & (42.2-78.9)  & $116 \pm  3  $ &         Arm II  & $ 18.3^{+ 0.4}_{-0.4}$  & (42.2-78.9) & $108 \pm  2 $ \\
\vspace{1 mm}
NGC 6384 Arm I*  & $ 24.5^{+ 0.4}_{-1.9}$ & (31.4-84.9)  & $125 \pm  6  $ &         Arm II  & $ 18.0^{+ 0.7}_{-1.6}$  & (40.1-57.2) & $ 63 \pm  4 $ \\
\vspace{1 mm}
NGC 7479 Arm I*  & $ 36.6^{+ 2.3}_{-1.4}$ & (56.1-95.8)  & $ 41 \pm  3  $ &         Arm II  & $ 23.0^{+ 2.8}_{-0.7}$  & (60.4-95.8) & $ 62 \pm  5 $ \\
\vspace{1 mm}
 IC 5325 Arm I   & $ 49.2^{+ 2.3}_{-9.7}$ & (11.4-23.0)  & $ 35 \pm  7  $ &         Arm II* & $ 20.3^{+ 3.0}_{-0.4}$  & (14.3-27.4) & $100 \pm  8 $ \\

\enddata

\tablecomments{
Column (1): object and spiral arm segment, see Figures~\ref{combo_1}-\ref{combo_4}. 
Columns (2) and (6): $H$-band pitch angles, $i_{H}$, in degrees.
Columns (3) and (7): radial ranges, $r_{0}$ to $r$, in arcsec.
Columns (4) and (8): azimuthal ranges , $\alpha = \cot{i_{H}} \ln{\left( \frac{r}{r_{0}} \right)}$, in degrees.
Column (5): spiral arm segment (see Figures~\ref{combo_1}-\ref{combo_4}) for the same object as Column (1).
}

\end{deluxetable}


\subsection{``Fourier Method''}~\label{sec_fourier}

\begin{figure*}
\centering
\includegraphics[scale=0.7]{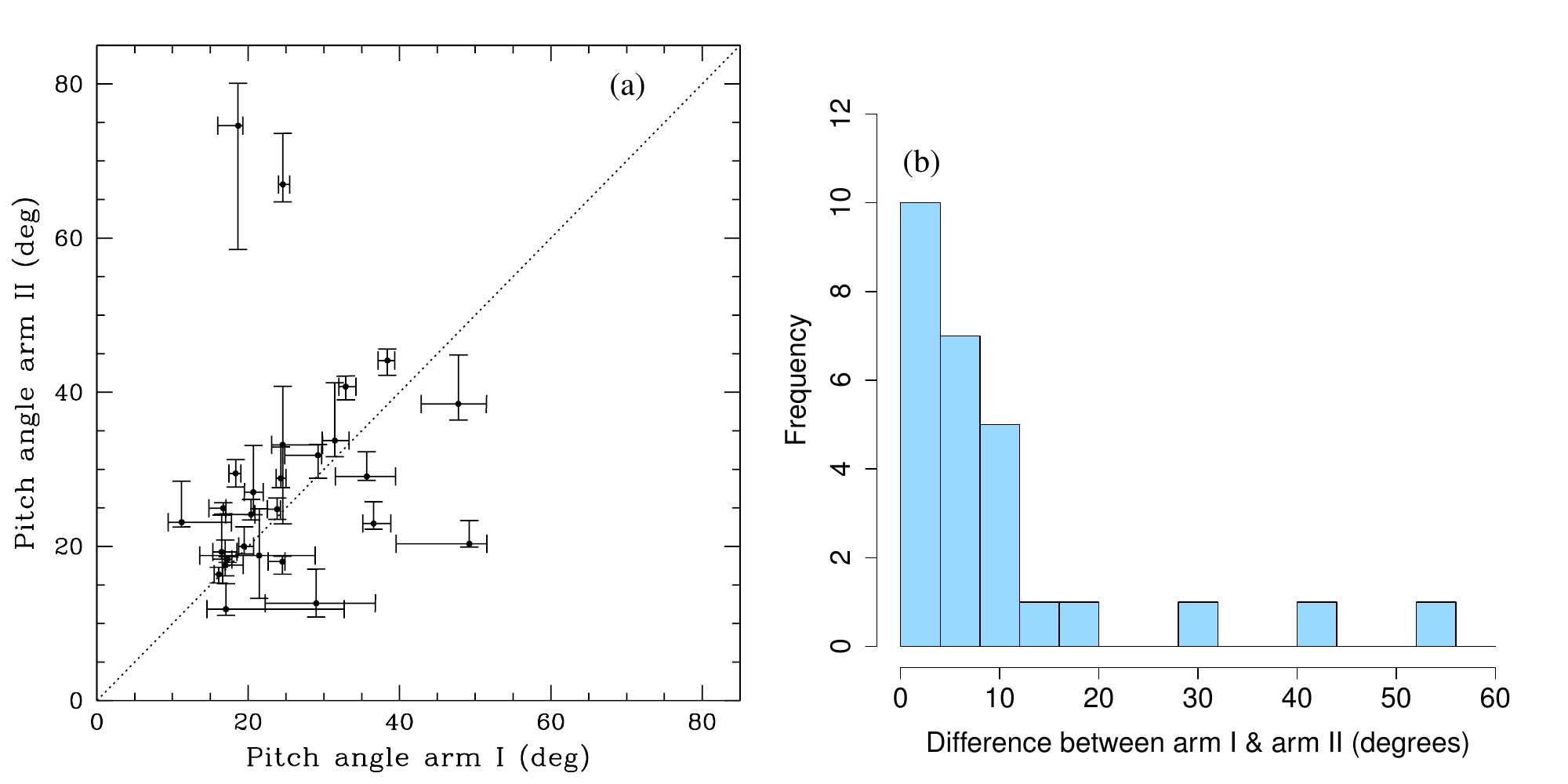}
 \caption{ (a) Pitch angles (in deg) for arm segments I ($x$-axis) vs.
arm segments II ($y$-axis) for each object.
Dotted line: one-to-one relation.
(b) Histogram of the absolute value difference between arm segments I and II,
obtained for each object with the ``slope method''. The standard deviation around the zero difference value
is $16\fdg3$.}
\label{fig_armIvsII}
\end{figure*}

Figure~\ref{fig_armIvsII}(a) plots the pitch angles in arm segment I
versus arm segment II for each object as obtained with the ``slope method''.
Figure~\ref{fig_armIvsII}(b) shows a histogram of the absolute value difference between arm segments I and II.
As shown in the figures, some scatter is present when analyzing spiral arm segments within the same galaxies.
Since we are interested in comparing single values of pitch angles for each object,
we need a method that provides the ``dominant mode'' for the pitch angle measurement.
The ``Fourier method'' is perfectly adequate for this purpose.

In this method, it is again assumed that the arms can be represented by logarithmic
spirals.\footnote{However, a Fourier analysis can be done without the assumption of a constant pitch angle.}
The Fourier amplitudes for each component are given by

\begin{equation}~\label{eq_Amp}
 A(m,p) = \frac{\sum_{i=1}^{I}\sum_{j=1}^{J} I_{ij}(\ln{r},\theta) \mathrm{exp}[-i(m\theta+p\ln{r})]}
          {\sum_{i=1}^{I}\sum_{j=1}^{J} I_{ij}(\ln{r},\theta)},
\end{equation}

\noindent where $r$ and $\theta$ are the polar coordinates, $I_{ij}$
is the intensity at coordinates $\ln{r}$, $\theta$, $m$ is the
number of spiral arms (or modes), and $p$ is related to the spiral
arms pitch angle ($i_{H}$) by

\begin{equation}
 \tan{i_{H}} = -m/p_{\mathrm{max}},
\end{equation}

\noindent where $p_{\mathrm{max}}$ corresponds to the maximum of $A(m,p)$
and $m=0,1,2,3,\dots$, i.e., the maximum of the Fourier spectrum
~\citep[see, e.g.,][]{pue92,sch94} for mode $m$.
Most of the analyzed objects present $m=2$ as the dominant mode for the spiral arms in the $H$-band
(see Table~\ref{tbl-FFT}), so it was adopted for this investigation.
The exceptions are NGC 3261 and NGC 4930 in which $m=1$ dominates and was used instead.
For NGC 1300 and NGC 7479, other Fourier modes ($m$) compete with the $m=2$ mode because
of the spiral arm segments with variable pitch angles. The pitch angles corresponding to the $m=2$ Fourier mode
were adopted for these objects in the subsequent analysis.

For the galaxies of the sample, it has been realized that the presence of foreground
stars does not affect the value of the pitch angle in general.
Nevertheless, caution must be taken when
foreground stars (or objects) compete in extension with spiral arms (see, e.g.,
annulus for NGC 864 in Figure~\ref{combo_1}). In these cases the need
for masks is required.
 
Objects were deprojected as explained in~Section~\ref{sec_slope}.
Radial ranges were selected to cover the spiral segments previously analyzed with the ``slope method''.
The azimuthal coverage is $2\pi$ radians.
The analyzed annuli are shown graphically in Figures~\ref{combo_1}-\ref{combo_4} (dashed lines).
These are the regions where the Fourier analysis was performed.

Table~\ref{tbl-FFT} shows the results for the Fourier pitch angle values,
which agree with the ``slope method'' within a $\sim 16\degr$ difference
(this corresponds to $1\sigma$ in Figure~\ref{fig_armIvsII}) in the majority of the objects.
NGC 5921 and NGC 6221 present the largest differences ($\sim 15\degr$).
For two objects, NGC 4995 and IC 5325, the computed pitch angles are close to $\sim 90\degr$.
This is due to the fact that the spiral arms have a low surface brightness (as compared to the disk)
and the bar component is difficult to isolate in the analyzed annulus. The ``slope method'', for the
``best-defined arm'', was used instead for these two objects in the subsequent analysis.

\subsubsection{Error Determination}
Errors were determined in the same way as in the ``slope method''.
These were added in quadrature with the error
intrinsic to the method. A program was built that computes the two-dimensional
fast Fourier transform in Equation~\ref{eq_Amp}. The output of this program is a $128 \times 2048$ ($m,p$) matrix.
The two closest values near $p_{\mathrm{max}}$ were used to approximate the error of the method.


\vspace{10 mm}

\section{COMMENTS FOR SOME OBJECTS}~\label{sec_OBJcomments}

{\it {NGC 210.}}
``Skinny'' spiral arms compared with the bar.

{\it {NGC 289.}}
The outer spiral arms have a greater pitch angle ($i_{H}\sim 40 \degr$)
as compared to the inner ones ($i_{H}\sim 25 \degr$; ``slope method'').

{\it {NGC 578.}}
Two symmetric spiral arms near the bar.

{\it {NGC 1073.}}
Spiral arms difficult to trace (low signal-to-noise ratio).

{\it {NGC 1187.}}
The two arm features analyzed are visually attached to the bar.
A third arm feature, not visually attached to the bar, is present.
The radial ranges for the ``Fourier method'' were modified with respect
to the ``slope method'' to allow a better signal-to-noise ratio in the
$\ln{r}$ versus $\theta$ map.

{\it {NGC 1300.}}
Two well-defined logarithmic spiral arms, although short in azimuthal range.
The adopted deprojection parameters were changed as compared to the
ones of~\citet{lau04}. This was done because the values provided in~\citet{lau04}
do not agree with the outer isophotes of the OSUBGS images.
An average between Hyperleda~\citep{pat03b}, RC3~\citep{deV91}, and a visual determination of the outer isophotes was used.

The deprojection parameters from~\citet{lin97}
were also tried for the pitch angle measurements.
These parameters, $\phi = 87 \degr\pm 2\degr $ and $q= 0.82 \pm 0.05$, are based on~\ion{H}{1} data
and are independent of kinematical or dynamical criteria~\citep[see also][]{kal10}.
Using these parameters, spiral arms are difficult to follow in a $\ln{r}$ versus $\theta$ map (assuming a logarithmic geometry).
For arm region I, a pitch angle of $21\fdg2  \pm 6\degr $ was obtained.
Arm region II was not possible to measure via the ``slope'' method.
The pitch angles obtained by applying the ``Fourier'' method
led to values with a contrary sign to the one expected, i.e., an inverse sense
of winding for the spiral arms.

{\it {NGC 1703.}}
Difficult to analyze the spiral arms in the inner
regions due to few pixels in a $\ln{r}$ versus $\theta$ map.

{\it {NGC 1832.}}
The bar region is distorted (not straight).

{\it {NGC 3059.}}
``Hard to follow'' logarithmic shape for the spiral arms.

{\it {NGC 3583.}}
Two symmetric spiral arms can be appreciated in the outer disk.
The region close to the bar presents a structure similar
to a ring or a tight spiral arm.

{\it {NGC 4145.}}
Double bar system?

{\it {NGC 4303.}}
This object presents three main spiral arms.

{\it {NGC 4902.}}
Three spiral regions are present in this object.

{\it {NGC 5921.}}
This object presents an inner ring and spiral features.

{\it {NGC 6300.}}
This object presents spiral features and apparently a ring feature.

{\it {NGC 6384.}}
Spiral arms with bifurcations. 

{\it {NGC 7479.}}
In general the spiral arms for this object do not present a clear 
logarithmic geometry.

{\it {IC 5325.}}
This object presents four well defined segments of spiral arms.
Only the ones near the bar's end were analyzed.



\begin{deluxetable}{lccccccllc}
\tabletypesize{\scriptsize}
\tablecaption{``Fourier Method" Derived Parameters~\label{tbl-FFT}}
\tablewidth{0pt}
\tablehead{
\colhead{Galaxy} & \colhead{$\frac{m_{1}}{m_{2}}$} & \colhead{$\frac{m_{2}}{m_{2}}$} & \colhead{$\frac{m_{3}}{m_{2}}$}
                 & \colhead{$\frac{m_{4}}{m_{2}}$} & \colhead{$\frac{m_{5}}{m_{2}}$} & \colhead{$\frac{m_{6}}{m_{2}}$}
                 & \colhead{$i_{H}$~(deg)} & \colhead{$i_{B}$~(deg)}  & \colhead{$\Delta{r}$~(arcsec)}
}

\startdata

\vspace{1 mm}
NGC 150  & 0.416 & 1.000 & 0.344 & 0.213 & 0.395 & 0.167 &  $ 27.9^{+0.9}_{-1.9}$    & $ 17.6^{+1.3}_{-2.0}$ & (34.7-55.4) \\
\vspace{1 mm}
NGC 210  & 0.525 & 1.000 & 0.189 & 0.478 & 0.255 & 0.220 &  $ 16.7^{+0.5}_{-0.5}$    & $ 15.7^{+0.5}_{-0.5}$ & (63.6-107.6)\\
\vspace{1 mm}
NGC 289  & 0.103 & 1.000 & 0.309 & 0.213 & 0.091 & 0.109 &  $ 19.7^{+0.8}_{-0.5}$    & $ 17.2^{+1.3}_{-1.8}$ & (22.8-30.8) \\
\vspace{1 mm}
NGC 578  & 0.656 & 1.000 & 0.289 & 0.255 & 0.080 & 0.116 &  $ 24.2^{+1.6}_{-1.5}$    & $ 23.0^{+0.9}_{-1.3}$ & (20.8-50.3) \\
\vspace{1 mm}
NGC 864  & 0.721 & 1.000 & 0.462 & 0.361 & 0.307 & 0.325 &  $ 20.7^{+1.2}_{-1.5}$    & $ 18.0^{+0.9}_{-0.9}$ & (33.0-68.5) \\
\vspace{1 mm}
NGC 1073 & 0.778 & 1.000 & 0.515 & 0.506 & 0.307 & 0.349 &  $ 34.7^{+1.4}_{-5.9}$    & $ 12.4^{+1.1}_{-0.5}$ & (37.8-70.0) \\
\vspace{1 mm}
NGC 1187 & 0.668 & 1.000 & 0.460 & 0.436 & 0.655 & 0.239 &  $ 21.4^{+1.3}_{-1.2}$    & $ 19.4^{+1.1}_{-1.0}$ & (29.0-59.4) \\
\vspace{1 mm}
NGC 1300 & 0.617 & 1.000 & 1.001 & 0.854 & 0.399 & 0.330 &  $11.2^{+15.3}_{-0.3}$    & $ 13.1^{+7.7}_{-0.3}$ & (74.3-116.9)\\
\vspace{1 mm}
NGC 1703 & 0.245 & 1.000 & 0.255 & 0.371 & 0.161 & 0.196 &  $ 18.8^{+1.5}_{-2.0}$    & $ 17.2^{+1.6}_{-0.6}$ & (15.6-28.1) \\
\vspace{1 mm}
NGC 1832 & 0.242 & 1.000 & 0.492 & 0.399 & 0.100 & 0.128 &  $ 25.1^{+1.8}_{-1.1}$    & $ 24.4^{+1.1}_{-1.6}$ & (20.0-45.8) \\
\vspace{1 mm}
NGC 3059 & 0.728 & 1.000 & 0.371 & 0.475 & 0.424 & 0.694 &  $ 27.0^{+3.0}_{-2.6}$    & $  8.9^{+0.5}_{-0.9}$ & (23.6-54.0) \\
\vspace{1 mm}
NGC 3261 & 1.043 & 1.000 & 0.246 & 0.226 & 0.259 & 0.322 &  $ 9.1^{+13.3}_{-0.3}$    & $10.2^{+20.1}_{-0.4}$ & (27.4-36.1) \\
\vspace{1 mm}
NGC 3513 & 0.481 & 1.000 & 0.229 & 0.472 & 0.141 & 0.220 &  $ 25.7^{+1.2}_{-1.7}$    & $ 24.2^{+1.7}_{-0.7}$ & (27.8-72.6) \\
\vspace{1 mm}
NGC 3583 & 0.446 & 1.000 & 0.470 & 0.225 & 0.300 & 0.146 &  $ 24.5^{+2.4}_{-2.1}$    & $ 22.6^{+2.0}_{-1.3}$ & (34.7-68.8) \\
\vspace{1 mm}
NGC 3686 & 0.450 & 1.000 & 0.886 & 0.271 & 0.276 & 0.493 &  $ 14.4^{+0.4}_{-0.6}$    & $ 14.4^{+0.6}_{-0.8}$ & (20.2-46.2) \\
\vspace{1 mm}
NGC 4145 & 0.448 & 1.000 & 0.250 & 0.348 & 0.240 & 0.275 &  $ 23.9^{+1.0}_{-0.7}$    & $ 17.3^{+0.4}_{-1.3}$ & (50.5-87.9) \\
\vspace{1 mm}
NGC 4303 & 0.260 & 1.000 & 0.174 & 0.294 & 0.149 & 0.186 &  $ 42.8^{+2.8}_{-1.9}$    & $ 37.7^{+1.6}_{-2.2}$ & (29.2-60.5) \\
\vspace{1 mm}
NGC 4902 & 0.467 & 1.000 & 0.210 & 0.665 & 0.277 & 0.401 &  $ 20.8^{+2.5}_{-1.6}$    & $ 25.3^{+3.5}_{-4.0}$ & (19.5-34.7) \\
\vspace{1 mm}
NGC 4930 & 1.081 & 1.000 & 0.550 & 0.272 & 0.181 & 0.223 &  $ 30.1^{+2.1}_{-2.1}$    & $ 13.9^{+0.9}_{-2.0}$ & (36.2-58.2) \\
\vspace{1 mm}
NGC 4995 & 0.898 & 1.000 & 0.356 & 0.595 & 0.234 & 0.204 &  $ 90.0^{+5.6}_{-8.8}$    & $ 78.3^{+5.4}_{-8.3}$ & (22.5-37.7) \\
\vspace{1 mm}
NGC 5483 & 0.163 & 1.000 & 0.134 & 0.439 & 0.158 & 0.280 &  $ 26.3^{+1.9}_{-1.8}$    & $ 21.6^{+1.4}_{-0.9}$ & (14.8-39.4) \\
\vspace{1 mm}
NGC 5921 & 0.188 & 1.000 & 0.454 & 0.524 & 0.152 & 0.313 &  $ 30.6^{+4.9}_{-3.1}$    & $ 22.5^{+2.0}_{-2.2}$ & (61.0-95.3) \\
\vspace{1 mm}
NGC 6221 & 0.602 & 1.000 & 0.315 & 0.413 & 0.115 & 0.235 &  $ 35.4^{+4.7}_{-1.4}$    & $ 19.9^{+1.6}_{-0.7}$ & (39.0-57.6) \\
\vspace{1 mm}
NGC 6300 & 0.199 & 1.000 & 0.201 & 0.330 & 0.157 & 0.283 &  $ 23.1^{+0.7}_{-0.9}$    & $ 20.3^{+0.6}_{-1.1}$ & (42.2-78.9) \\
\vspace{1 mm}
NGC 6384 & 0.501 & 1.000 & 0.574 & 0.312 & 0.385 & 0.273 &  $ 26.3^{+1.2}_{-1.2}$    & $ 28.0^{+0.9}_{-2.0}$ & (31.4-84.9) \\
\vspace{1 mm}
NGC 7479 & 1.073 & 1.000 & 0.656 & 1.200 & 0.610 & 0.473 &  $ 26.8^{+3.8}_{-1.8}$    & $ 31.5^{+1.6}_{-3.2}$ & (56.1-95.8) \\
\vspace{1 mm}
 IC 5325 & 0.367 & 1.000 & 0.426 & 0.176 & 0.200 & 0.090 &  $ 69.4^{+3.8}_{-5.2}$    & $81.5^{+4.2}_{-12.7}$ & (11.4-27.4) \\

\enddata

\tablecomments{
Column (1): galaxy name. 
Column (2): ratio between the maximum amplitudes of Fourier modes $m=1$ and $m=2$, in the $H$-band. 
Columns (3), (4), (5), (6) and (7): ratio between the maximum amplitudes of the respective Fourier modes,
in the $H$-band. 
Column (8): $H$-band (see~Section~\ref{sec_fourier}) pitch angles, in degrees.
Column (9): $B$-band (see~Section~\ref{denw_pred}) pitch angles, in degrees.
Column (10): radial ranges, $r_{0}$ to $r$, in arcsec.
}

\end{deluxetable}


\section{RESULTS AND DISCUSSION}~\label{sec_resdis}

\begin{figure}
\centering
\epsscale{1.05}
\plotone{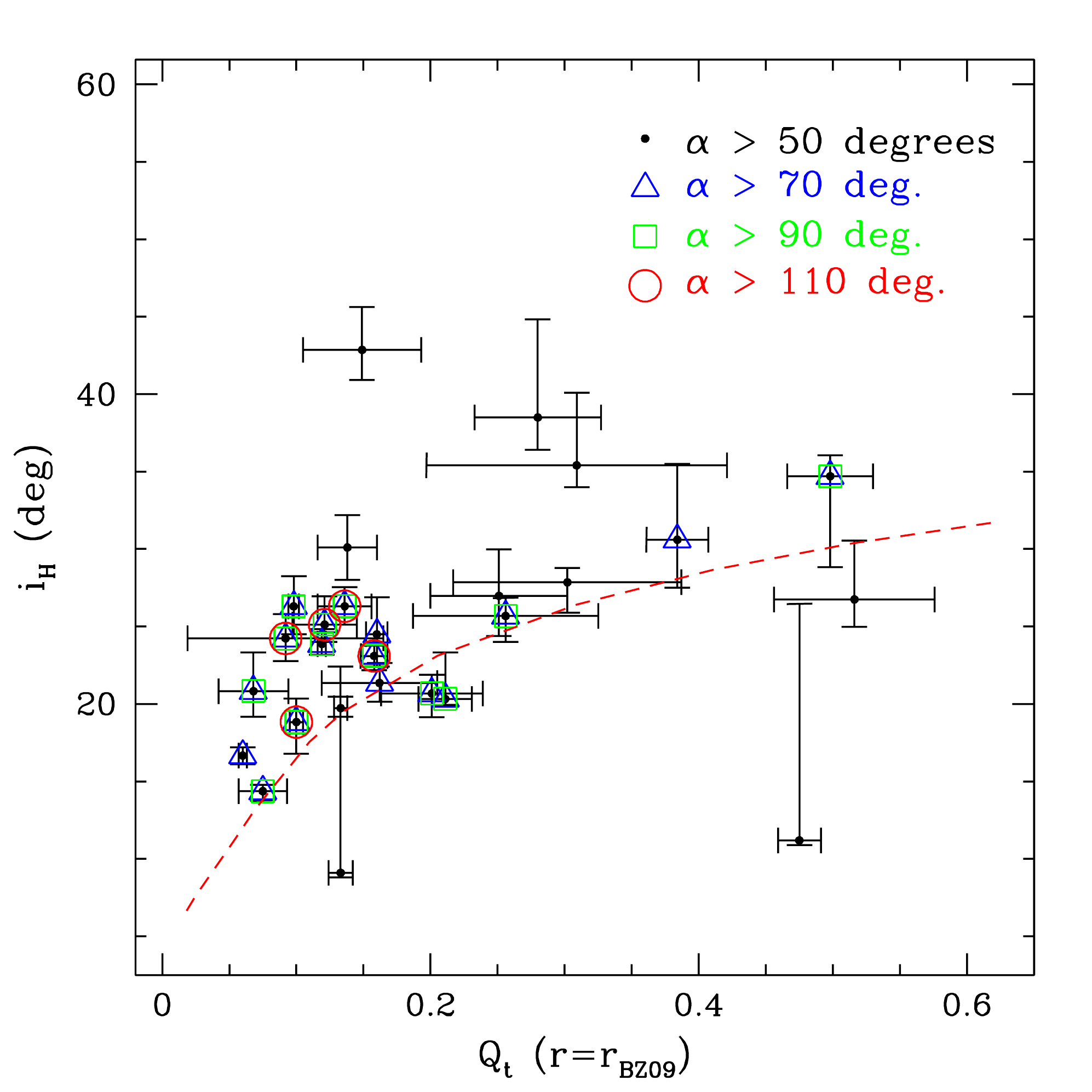}

 \caption{Pitch angle $i_{H}$ in deg
vs. perturbation strength $Q_{t}(r=r_{\mathrm{BZ09}})$
for the 27 galaxies selected for analysis (see~Section~\ref{gal_sample}).
Lagrangian radius, $r_{L} = r_{\mathrm{BZ09}}$, from~\citet{butz09}.
The dashed line corresponds to the predicted correlation
for models A and D in~\citet{atha09b}. Data are separated by $\alpha > 50 \degr$ (all points),
$\alpha > 70 \degr$ (triangles), $\alpha > 90 \degr$ (squares), and $\alpha > 110 \degr$ (circles).
}
\label{fig_QtL1}
\end{figure}

Figure~\ref{fig_QtL1} shows the results for the pitch angle, $i_{H}$ (Fourier method, except for NGC 4995,
and IC 5325, see~Section~\ref{sec_fourier}), versus
perturbation strengths, $Q_{t}(r=r_{\mathrm{BZ09}})$. A first inspection of the data,
where the ``azimuthal range''\footnote{This is obtained with the ``slope method'' via Equation~\ref{eq_alpha}.
It is the ``maximum'' azimuthal range that is taken into account, i.e., the greatest value of $\alpha$ for either arm segment I or II.}
is $\alpha > 50\degr$, shows considerable scatter
around the predicted correlation for models A~\citep[][bar potential]{fer77}
and D~\citep[][bar potential]{deh00} in~\citet{atha09b}.\footnote{A third model with a~\citet{bar67}
bar potential (BW model) was considered in~\citet{atha09a,atha09b}. This model agrees with model D
up to $Q_{t}(r=r_{L}) \sim 0.2$, and deviates toward higher pitch angles
afterward, up to $\sim 5 \degr$ at $Q_{t}(r=r_{L}) \sim 0.6$.}
However, if the $\alpha$ criterion is changed to logarithmic spiral segments that extend up to
$\alpha > 70\degr$, $\alpha > 90\degr$, and $\alpha > 110\degr$, the scatter
is reduced. The reduced Pearson's chi-square, $\chi^{2}/n$, obtained as
\begin{equation}
 \chi^2 = \sum_{k=1}^{n} \frac{(i_k - i_p)^2}{i_p},
\end{equation}

\noindent where $i_{k}$ is the $k$th Fourier-measured pitch angle
and $i_p$ is the predicted pitch angle value for models A and D in~\citet{atha09b},
gives the results 3.10, 1.55, 1.83, and 2.00 for $\alpha>50\degr$ ($n=27$), $\alpha>70\degr$ ($n=17$),
$\alpha>90\degr$ ($n=13$), and $\alpha>110\degr$ ($n=5$), respectively. 

\begin{figure}
\centering
\epsscale{1.05}
\plotone{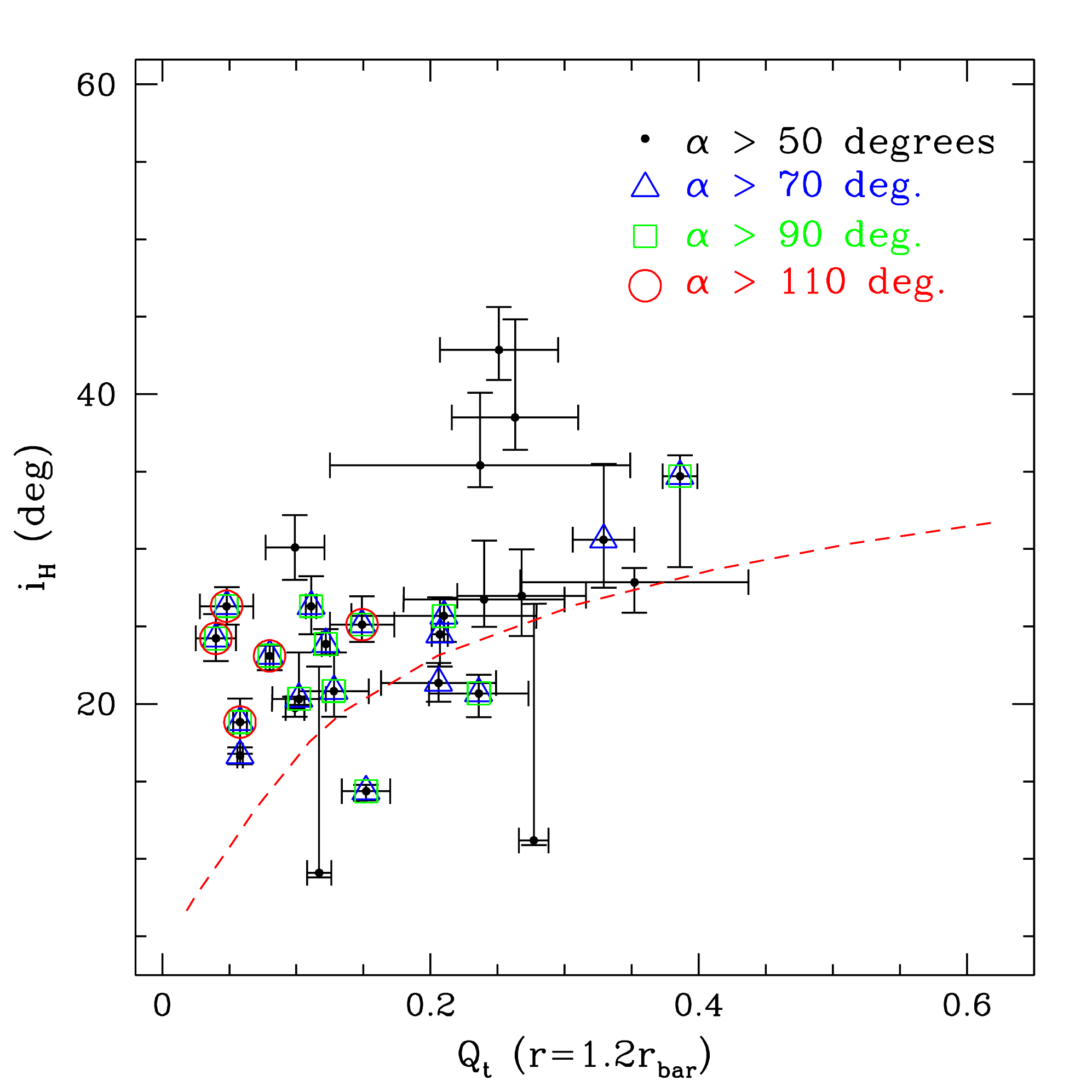}

 \caption{Same as Figure~\ref{fig_QtL1} for perturbation strength $Q_{t}(r=1.2 r_{\mathrm{bar}})$.}
\label{fig_Qt12Rbar}
\end{figure}

\begin{figure}
\centering
\epsscale{1.05}
\plotone{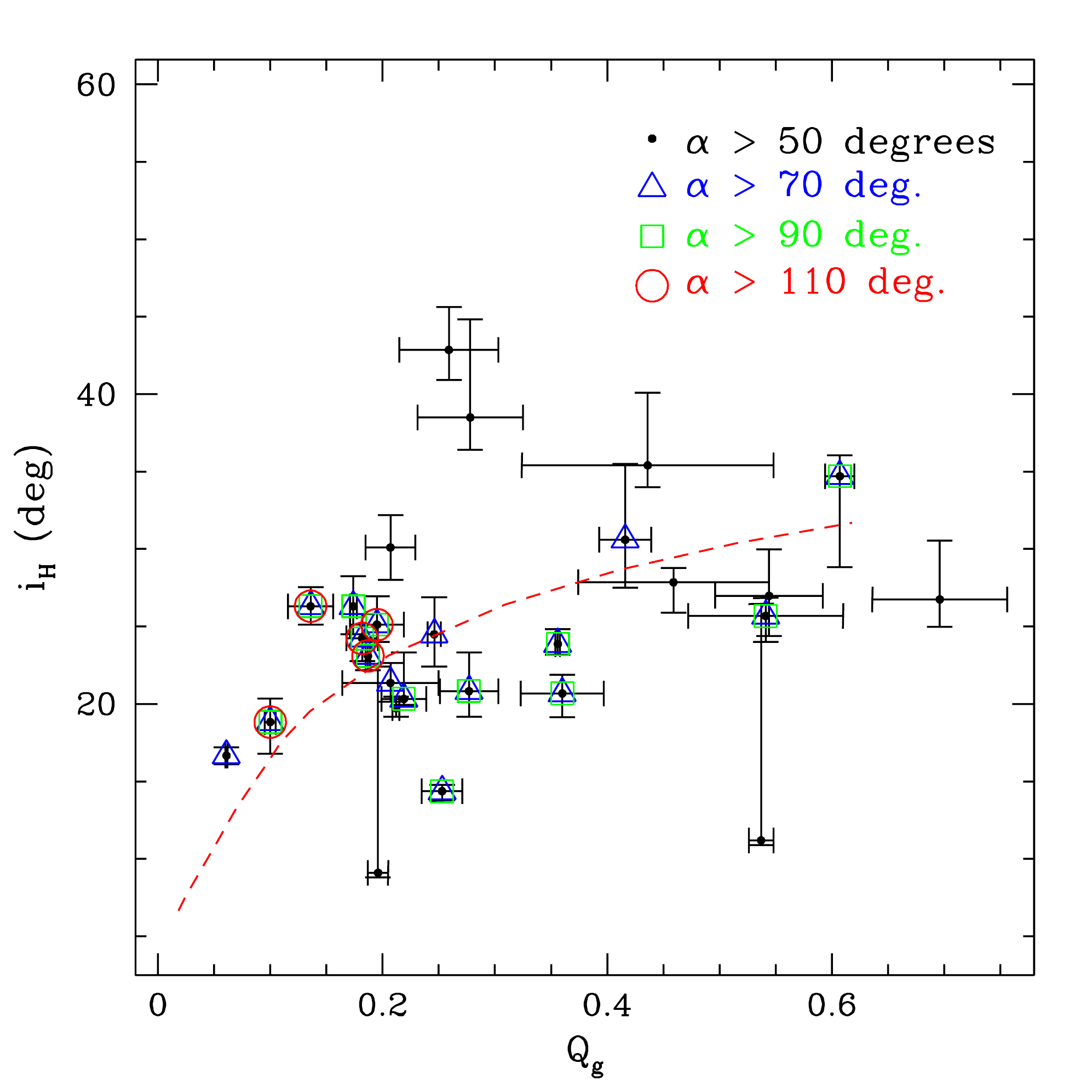}

 \caption{Same as Figure~\ref{fig_QtL1} for perturbation strength $Q_{g}$.}
\label{fig_Qg}
\end{figure}

Figures~\ref{fig_Qt12Rbar} and~\ref{fig_Qg} show the results for the cases
$Q_{t}(r=1.2r_{\mathrm{bar}})$ and $Q_{g}$, respectively. For $\alpha > 70\degr$,
reduced Pearson's chi-square values obtained as
\begin{equation}
 \chi^2 = \sum_{k=1}^{n} \frac{(Q_k - Q_p)^2}{Q_p},
\end{equation}

\noindent where $Q_{k}$ is the $k$th bar strength value corresponding
to the $k$th Fourier-measured pitch angle
and $Q_p$ is the predicted bar strength value for models A and D in~\citet{atha09b},
yield the results 0.049, 0.075, and 0.084
for the $Q_{t}(r=r_{\mathrm{BZ09}})$, $Q_{t}(r=1.2r_{\mathrm{bar}})$, and $Q_{g}$
plots, respectively.

According to this result, the best concordance with the~\citet{atha09b} model is obtained
by comparing the pitch angles with $Q_{t}(r)$ given at~\citet{butz09} bar corotation radii
($r=r_{\mathrm{BZ09}}$). This last point is not discussed in~\citet{atha09b}.

One important aspect in the~\citet{atha09b} prediction is that the self-gravity
of the spirals was not taken into account. The potential created by the ``confined''
chaotic orbits is neglected. Contrarily,~\citet{tso09} emphasize the
contribution of the spiral part for studying the dynamics of the ``chaotic''
spirals.
Also, realistic bar potentials are hard to model.
If many different realistic potentials are used, the predicted correlations
may become broader~\citep{atha09b,atha10}. This may explain in
Figure~\ref{fig_QtL1} the tendency of the points (squares and circles) 
to be above the predicted correlation for $Q_{t}(r=r_{\mathrm{BZ09}}) < 0.2$.


\subsection{Density Wave Theory Prediction}~\label{denw_pred}

The modal approach explains the density wave phenomena
as generated by intrinsic mechanisms in the disk~\citep{bert89a,bert89b,ber96}.
Normal modes of oscillation generate spontaneously and evolve according to
the physical and dynamic properties of the system. Three physical properties
determine the morphology in disk galaxies: the disk mass, the gas content,
and the stellar velocity dispersion. When the disk mass is ``high", bar structures
are generated as oscillating modes of the system. The modal theory considers
bars and spirals equally, i.e., as normal modes of oscillation in the disk.

Based on the dispersion relation, linear density wave theory predicts~\citep{hoz03}
that the pitch angle should increase with increasing velocity dispersion, or that

\begin{equation}~\label{eq_tani}
   \tan{i} \propto c^{2}_{r}/\Sigma,
\end{equation}

\noindent where $i$ is the arms pitch angle, $c_{r}$ is the radial
velocity dispersion, and $\Sigma$ is the surface density of the disk.
Spiral structure shows different morphologies when observed in optical versus NIR
bands~\citep{blo91,gros98}. NIR bands can trace both the old populations of bar and spiral arms, 
assuming that red young stars do not contribute globally to the observed radiation~\citep{rix93}.
Also, older populations have a higher velocity dispersion compared to younger
ones~\citep{bar67,wie77,nor04,bin08}.
For most galaxies at the arm location, we have that $\sim 98\%$ (by mass) of the stars belong to
evolved populations~\citep[see][and references therein]{gon96}. Nevertheless, young
stars contribute to most of the light in optical wavelengths.
According to this, NIR images of spiral perturbations should present higher
pitch angles compared to optical ones.
Azimuthal age (color) gradients~\citep[e.g., ][]{gon96,mar09a,mar09b,mar11} may also
affect the pitch angles observed in the optical versus NIR bands, but these are very
difficult to trace by just comparing the light distributions in two
bands~\citep{gon96,sei98}. Besides, azimuthal gradients are not located
continuously along the spiral arms but in specific regions~\citep{gon96,mar09a,mar11}.

From Equation~\ref{eq_tani}, taking into account that young and old stars
are similarly affected by the gravitational potential of the disk (which depends on the surface density),
we obtain

\begin{equation}~\label{eq_iB}
   i_{B} = \arctan { \left\{ \left( \frac{c_{r_{B}}}{c_{r_{H}}} \right)^{2} \tan{i_{H}} \right\} },
\end{equation}

\noindent where $i_{B}$ is the $B$-band pitch angle, $i_{H}$ is the $H$-band pitch angle,
$c_{r_{B}}$ is the radial velocity dispersion of young stars, and $c_{r_{H}}$ is the radial velocity
dispersion of old stars.

In the case of the invariant manifold theory, where chaotic orbits are ``confined'' in the
spiral locus, no difference between pitch angles of spiral arms traced in different
wavelengths is predicted~\citep{atha10}.

\begin{figure}
\centering
\epsscale{1.05}
\plotone{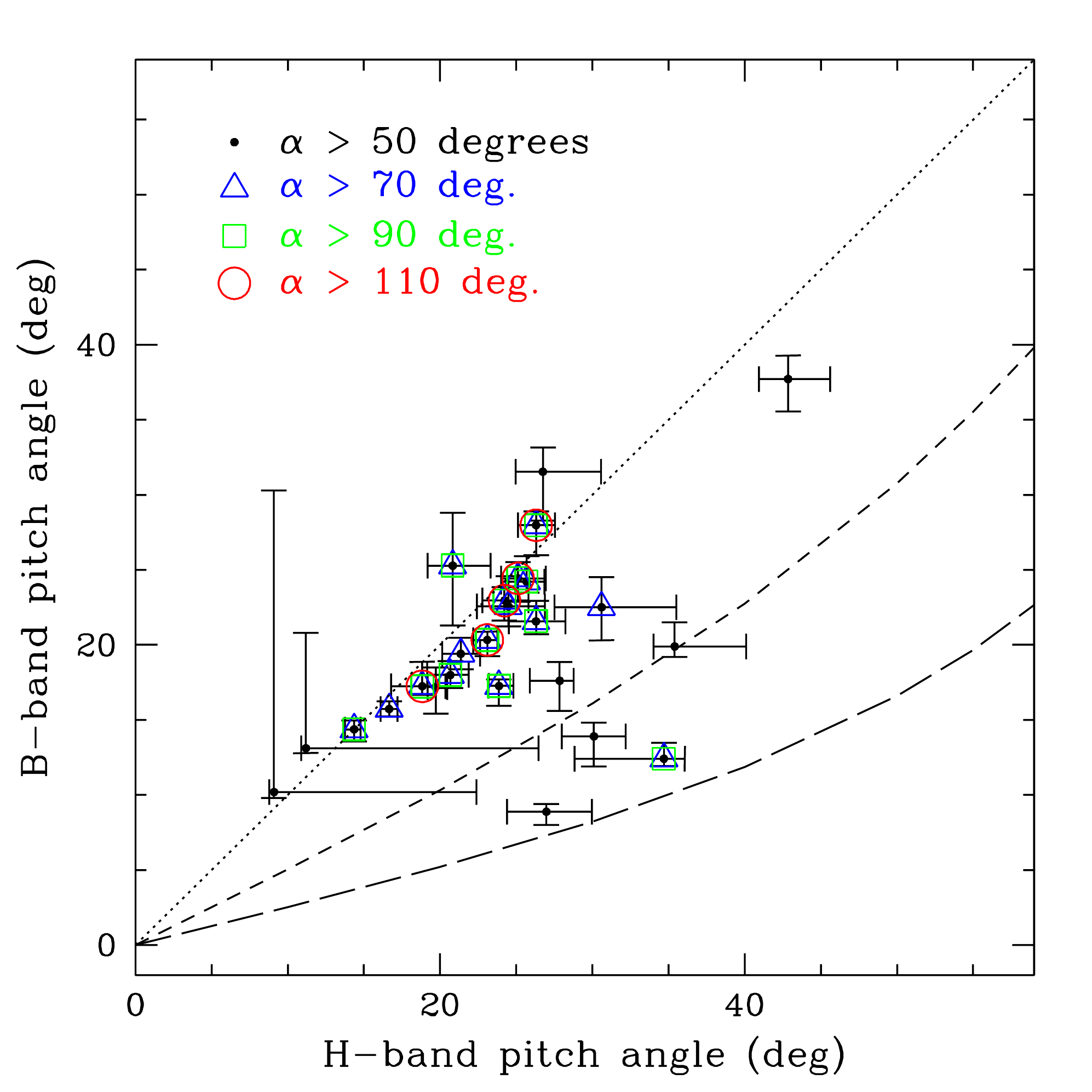}

 \caption{$H$-band pitch angle vs. $B$-band pitch angle, obtained via the ``Fourier'' method.
Dotted line: one-to-one relation;
short-dashed line: density wave theory prediction from Equation~\ref{eq_iB},
assuming $c_{r_{H}} \sim \sqrt{2}  c_{r_{B}}$;
long-dashed line: density wave theory prediction from Equation~\ref{eq_iB},
assuming $c_{r_{H}} \sim 2 c_{r_{B}}$. }
\label{fig_HandB}
\end{figure}

\citet{sei06} found a nearly 1:1 correlation between pitch angle measurements 
in the $B$ and $H$ bands, for 57 galaxies in the OSUBGS~\citep{esk02} sample.
Nevertheless, based on the sample of five non-barred and weakly barred spirals,
~\citet{gros98} notice that the main two-armed spiral is tighter when measured in bluer colors.
For the barred-spirals data presented in this investigation, we measured the pitch angles in the $B$-band images
for the same objects analyzed in the $H$-band from the OSUBGS sample, applying the ``Fourier'' method.
The $B$-band images were registered to the $H$-band images, so the
high-resolution data ($B$-band) were degraded to the low-resolution data ($H$-band in this case).
Annulus regions were selected in the same positions as the
$H$-band, and the pitch angles were measured identically\footnote{The same
dominant modes, $m=1$ or $m=2$ (see~Section~\ref{sec_fourier}), as measured in the $H$-band were
adopted for the $B$-band pitch angle measurements.}
with the method described in~Section~\ref{sec_fourier}.
The results are shown\footnote{NGC 4995 and IC 5325 were excluded from this analysis (see the last paragraph in~Section~\ref{sec_fourier}).}
in Figure~\ref{fig_HandB} (and Table~\ref{tbl-FFT}) where a tendency of $\sim 30\%$ of the points
toward higher $H$-band pitch angles is observed. Although, if we apply
the same azimuthal range ($\alpha$) criteria as in Figures~\ref{fig_QtL1},~\ref{fig_Qt12Rbar}, and~\ref{fig_Qg},
we can notice that $\sim 80\%$ of the $\alpha > 70 \degr$ data lie very close to the 1:1 relation as expected (independently of $\alpha$)
from~\citet{atha10}.


\subsection{Invariant Manifolds as Apsidal Sections}

A comparison of the different treatments of the manifolds
viewed as apsidal sections~\citep{vog06b,vog06c,tso08,tso09}, or as tubes
that guide chaotic orbits~\citep{rom06,rom07,atha09a,atha09b,atha10},
requires a different analysis involving separating spirals and bars.
This will be covered in a subsequent publication.

\section{CONCLUSIONS}

The results of this investigation show the following.

\begin{enumerate}

\item Although the adopted deprojection parameters
may introduce some biases~\citep[see, e.g.,][]{bar03},
a trend can be observed where some strong barred spirals have more open spiral arms
when compared to galaxies with weaker bars.
This kind of trend was also discussed in~\citet{blo04}, where a similar behavior was found.
The correlation predicted by the manifold models of~\citet{rom06,rom07} and~\citet{atha09a,atha09b,atha10}
is better reproduced by observations on two conditions.
\begin{enumerate}
\item The corotation values obtained with the ``potential-density phase shift method''~\citep{butz09} are adopted.
\item The spirals logarithmic geometry is maintained for large azimuthal ranges, $\alpha > 70\degr$.
\end{enumerate}

\item The $\sim 60\%$ of the 27 galaxies on the analyzed sample
seem to reproduce the investigated correlation.

\item The pitch angles calculated via the ``Fourier method'' in the $B$ (young stars) and the $H$ (mostly old stars)
bands yield similar values for $\sim 80\%$ of the objects where the azimuthal range, $\alpha$, is greater than $70\degr$.
This kind of behavior is expected in the ``Lyapunov tube model''~\citep{atha10}, although
no restriction on the azimuthal range was given by the authors.

\item Other possible mechanisms to generate spiral features in barred galaxies,
such as bar-driven spirals~\citep[e.g.][]{sal10}, models where the Lagrangian
points of the system are specified by both bar and spirals~\citep[e.g.,][]{tso09},
or chaotic spirals inside corotation~\citep[thus not related with the presence of unstable
Lagrangian points;][]{pat10}, cannot be excluded by the present investigation.

\end{enumerate}

\section*{Acknowledgments}

I am grateful to the anonymous referee for many important remarks and helpful comments
that have improved this paper.
I acknowledge postdoctoral financial support from UNAM (DGAPA), M\'exico.
I thank Christos Efthymiopoulos for clarifying my inquiries about spiral arms driven by ``manifolds''.
This work made use of data from the Ohio State University Bright Spiral Galaxy Survey, which was funded by grants AST-9217716 and AST-9617006 from the United States National Science Foundation,
with additional support from the Ohio State University.


\end{document}